\documentclass[prb,showpacs]{revtex4}

\usepackage{amsmath}
\usepackage{amsfonts}
\usepackage{amssymb}
\usepackage{array}
\usepackage{floatflt}
\usepackage{graphicx}

\begin{document}

%\preprint{}

%%% to see comments, use first version, to hide them, use second one
\newcommand{\stef}[2]{$\blacktriangleright${\sc round #1:} {\em #2}$\blacktriangleleft$}

\title{Charge-statistics separation and probing non-Abelian states.}

\author{D. E. Feldman and Feifei Li}
\affiliation{Department of Physics, Brown University,
Providence, RI 02912, USA}

\begin{abstract}

Several states were proposed as candidates for the $\nu=5/2$ quantum Hall plateau. We suggest an experiment which can determine the physical state.
The proposal involves transport measurements in the geometry with three quantum Hall edges connected by two quantum point contacts.
In contrast to interference experiments, this approach can distinguish the Pfaffian and anti-Pfaffian states
as well as different states with identical Pfaffian or anti-Pfaffian statistics.
The transport is not sensitive to the fluctuations of the number of the quasiparticles trapped in the system.

\end{abstract}


\maketitle

%\section{Introduction}

The Pauli spin-statistics theorem does not apply in two dimensions and anyons can exist in addition to usual fermions and bosons. Different types of anyonic statistics can be classified as Abelian and non-Abelian.
In the former case, the wave function of an anyon system acquires a phase factor after one particle makes a circle around other anyons. In the latter case not only the wave function but also the quantum state changes after a particle moves along a closed loop.

Gauge invariance guarantees that fractionally charged excitations of electronic systems must be anyons. Hence, the observation of fractional charges in quantum Hall systems \cite{weizmann,Grenoble,goldman} proves the existence of fractional statistics. However, the experiments \cite{weizmann,Grenoble,goldman,goldman2005} have provided little information about the details of the quasiparticle statistics in the quantum Hall effect (QHE). In particular, the existence of non-Abelian anyons \cite{review} remains an open problem.

A promising place to look for non-Abelian statistics is the QHE plateau at the filling factor $\nu=5/2$. The nature of the 5/2 QHE state has not been understood yet and most theoretical proposals involve non-Abelian statistics \cite{MR,anti,rec}. However, a simpler Abelian state is also a possibility \cite{ab,rec}.
Six simplest theoretical proposals are described in Table I (see also Ref. \onlinecite{rec}).

\begin{table}%[h]
\vspace{3mm}
\begin{tabular}{|c|c|c|c|}
\hline
state & modes & statistics & signature \\
\hline
$K=8$ & 1R & A & $F_2=4e;$ $I_1,S_1=0$\\
Pfaffian & 3/2R & N & $F_2=5/2e;$ $I_1,S_1=0$\\
331 & 2R& A & non-universal $F_2$; $I_1,S_1=0$\\
edge-reconstructed
Pfaffian & 2R+1/2L& N & $S_1\ne0, I_1=0$\\
non-equilibrated anti-Pfaffian & 1R+3/2L& N & $F_1$ independent of the edge shape\\
disorder-dominated anti-Pfaffian & 1R+3/2L& N & $F_1$ depends on the edge shape\\
\hline
\end{tabular}
\caption{Proposed 5/2 states. The 2nd column shows the numbers of the right- and left-moving modes (R and L), Majorana fermions being counted as 1/2 of a mode.
'A' and 'N' denote Abelian and non-Abelian statistics.}
\label{tableI}
\vspace{3mm}
\end{table}

Several interferometry experiments were proposed \cite{FP,mz} for probing the 5/2 state.
The simplest approach is based on the Fabry-Perot geometry \cite{FP}. This approach can however work only if the number of the quasiparticles
 trapped in localized states inside the interferometer does not fluctuate on the time scale of the experiment \cite{kane}.
Since the energy gap for neutral excitations is likely to be low, this condition may not be easy to satisfy at realistic temperatures.
 The Mach-Zehnder interferometry is free from this limitation but shares another limitation with the Fabry-Perot approach:
it cannot distinguish any of the non-Abelian states listed in Table I from each other. So far interferometry \cite{new-q4} and other approaches \cite{q4,q41}
allowed the measurement of the quasiparticle charge $q=e/4$ in the 5/2 state. This is not sufficient for the determination of the physical 5/2 state since
$q=e/4$  is predicted by all theories of the 5/2 plateau listed in Table I.

Thus, some method other than interferometry is desirable. One idea consists in checking scaling relations \cite{rec,fiete} such as the power dependence of the current on the voltage $I\sim V^s$ in a quantum point contact \cite{rec}. Unfortunately,
even in the simplest case of the Laughlin states the theory has not been reconciled with the measurements of the $I-V$ curve \cite{chang}.
Besides, this approach is not expected to distinguish the anti-Pfaffian and edge-reconstructed Pfaffian states \cite{rec}.

In this paper we suggest another approach which leads to qualitatively different
results for all states listed in Table I and is not sensitive to the number of the trapped quasiparticles. It involves transport measurements in the geometry
illustrated in Fig. 1. This geometry is similar to the one used in experiments with Laughlin states \cite{bunching}.
Edge 3 connects sourse S3 with drain D3 and 
separates regions with filling factors 5/2 and 2. Edges 1 and 2 
connect source S1 with drain D1 and source S2 with drain D2 and
separate the region with filling factor 2 from the region
with filling factor 0. Electrons can tunnel across the integer QHE region through quantum point contacts QPC1 and QPC2
with the tunneling amplitudes $\Gamma_1$ and $\Gamma_2$ (Fig. 1), respectively,
at the distance $a$ from each other. Source S3 is maintained at zero voltage.
We consider two situations for the electric potentials of sources S1 and S2: 1) $V(S1)=V$, $V(S2)=0$ and 2) $V(S1)=0$, $V(S2)=V$.
In the first case we calculate the current $I_2$ and noise $S_2$ in drain D2. In the second case we find the current $I_1$ and noise $S_1$ in drain D1.
Nonzero $I_1$ is possible only in some states due to the presence of contra-propagating edge modes on the boundary between the $\nu=5/2$ and $\nu=2$ regions.
 We will see below that different states can be distinguished by zero/nonzero $I_1$ and/or $S_1$ and universal or non-universal Fano factors
$F_1=S_1/I_1$ and $F_2=S_2/I_2$. The outcomes of the proposed experiment for different states are summarized in Table I.
The physics is analogous in a similar geometry with all three edges separating a QHE liquid with $\nu=5/2$ from regions with $\nu=0$.
 However, an analytic calculation is impossible in that setup and its numerical analysis will be discussed elsewhere.

The system in Fig. 1 has the following Lagrangian
\begin{equation}
\label{1}
L=\sum_{k=1}^3L_k -\int dt \sum_{k=1}^2 [\Gamma_k T_k+h.c.],
\end{equation}
where $L_k$ are the Lagrangians of the three edges, $\Gamma_k$ the tunneling amplitudes at the two QPC's, $T_k=\epsilon_k^\dagger \psi(x_k)\eta_k$ the tunneling operators, $\epsilon_k$ the electron annihilation operators
on edges 1 and 2, $\psi(x_k)$ the electron annihilation operators on edge 3 ($x_1=0$, $x_2=a$) and the Klein factors $\eta_k$ make sure that the tunneling operators
$T_k$ commute. The operator $\psi(x)$ and  action $L_3$ of fractional QHE edge 3 depend on the model (Table I). $L_{1,2}$ describe chiral Fermi-liquid systems.
Edges 1 and 2 have two channels with spin up and down but it is sufficient to include only one of them in $L_{1,2}$ because of the spin conservation at the tunneling events.
The zero-temperature correlation functions of the fields $\epsilon$ assume the Fermi-liquid form
$\langle\epsilon_k^\dagger(t_1)\epsilon_k(t_2)\rangle\sim 1/[i(t_1-t_2)+\delta]$.

Below we use the perturbation theory to calculate the current and noise in the order $\Gamma_1^2\Gamma_2^2$. 
The perturbative calculation is legitimate if
$\Gamma^2(eV)^{2s+1}/E_c^{2s+2}<eV$, where $s$ is the scaling dimension of the operators $T_k$ and $E_c$ 
the cut-off energy of the order of the QHE energy gap.
We neglect the thermal noise $\sim\Gamma^2T^{2s+1}$. For  $\Gamma\sim E_c^{s+1}/(eV)^s$ this condition reduces to $eV>T$. 
Thus, we concentrate on the low-temperature limit $T=0$. We assume that the distance $a$ between the point contacts 
is sufficiently large, $a\gg a_V=hv/(eV)$, where $v$ is of the order of 
the edge mode velocity. This will allow us to treat QPC1 and QPC2 as independent.
At the same time we neglect equilibration between different co- and contra-propagating edge modes on the scale $a$. 
Indeed, if the Lagrangian of any model in Table I contains a large contribution, responsible for edge equilibration, 
or such contribution renormalizes to a large value at the scale $a_V$ then 
it opens a gap and/or modifies the character of soft modes and hence changes the model.

We begin with the simplest case of the $K=8$ state \cite{ab,rec}. Calculations are similar
but longer for the other models. The simplifications is due to the existence of only one edge mode on edge 3 
in the $K=8$ model whose Lagrangian
$L_3/\hbar=-(2/\pi)\int dtdx[\partial_t\phi\partial_x\phi+v(\partial_x\phi)^2]$, the charge density on the edge being $e\partial_x\phi/\pi$.
 In contrast to all other states in Table I, only electron pairs but not electrons can tunnel into the $K=8$
edge \cite{f0}.
Thus, $\psi=\exp(8i\phi)$ and $\epsilon_k$ should be understood as pair annihilation operators, $\langle\epsilon^\dagger_k(t)\epsilon_k(0)\rangle\sim 1/(it+\delta)^4$.

The operator of the current through QPC2 can be found as the time derivative of
the charge on edge 2, $I_2=(iq/\hbar)[\Gamma_2^* T_2^\dagger-\Gamma_2 T_2]$, where $q$ is the carrier charge. 
The noise $S_2=2\int_{-\infty}^0\langle I_2(t)I_2(0)+I_2(0)I_2(t)\rangle dt$. We assume that $qV(S_1)<0$, i.e., 
the chemical potential of edge 1 is lower than the potentials of edges 2 and 3. It is convenient to switch to the interaction representation such 
that all three chemical potentials become 0. This introduces a time-dependence into the operator
$T_1\sim \exp(iqVt/\hbar)$. We use the 4th order Keldysh perturbation theory.
Every nonzero contribution to $I_2$ and $S_2$ includes all four operators $T_1, T_1^\dagger, T_2, T_2^\dagger$. One finds

\begin{equation}
\label{2}
I_2=q(i/\hbar)^4|\Gamma_1\Gamma_2|^2\int d\tau d\Delta t dt T_c\langle T_1(\tau)T_1^\dagger(\tau+\Delta t)[T_2(t)T_2^\dagger(0)-T_2^\dagger(t)T_2(0)]\rangle,
\end{equation}
where the integration extends over the Keldysh contour, Fig. 2, $T_c$ denotes Keldysh time-ordering.
The time scale $t_V\sim a_V/v$ is set by the voltage bias $V$. The main contribution to the integral comes from small $t\sim t_V$ and $\Delta t\sim t_V$ and from large
$\tau {\approx} {-} {a/v}$.
This allows us to extend the integration with respect to
$\Delta t$ and $\tau$ from $-\infty$ to $+\infty$ on both upper and lower branches of the Keldysh contour (Fig. 2) 
keeping still the same ordering of operatros in Eq. (\ref{2}) as at negative $\tau\gg t,\Delta t$. Technically this corresponds to
shifting a variable, $\tau\rightarrow\tau+a/v$, and taking the limit $a\rightarrow\infty$.
The result of the integration with respect to $\tau$ is nonzero only if $T_1$ and $T_1^\dagger$ are located on different branches. The integration with respect to $\Delta t$ yields a nonzero result only if $T_1^\dagger$ is on the upper (i.e., later) branch. Finally, it is convenient to group together

1) The first term from the square brackets in Eq. (\ref{2}) with
$t$ on the upper branch and the second term with $t$ on the lower branch; and

2) The first term from the square brackets with $t$ on the lower branch and the second term with $t$ on the upper branch.

One obtains two contributions $I^{\pm}$ to $I_2$, proportional to the following integrals:

\begin{equation}
\label{3}
I^{\pm}=\int dt d\Delta t d\tau \frac{\exp(i|qV|\Delta t/\hbar)}{(\delta+i\Delta t)^{12}}
\frac{1}{(\delta\pm it)^{12}}\frac{[\delta+i(\tau+\Delta t)]^8[\delta+i(t-\tau)]^8}{[\delta+i(\tau+\Delta t-t)]^8[\delta-i\tau]^8}.
\end{equation}
The $t$ integration in $I^-$ yields 0 since both poles are in the lower half-plane. Thus, the current is proportional to $I^+$. 
This integral can be calculated analytically but its value is unimportant for the calculation of the Fano factor $F_2=S_2/I_2$. 
One can easily check that $S_2$ reduces to the same integral $I^+$ and $F_2=2q=4e$. The result is easy to understand. 
It reflects the fact that the current $I_2$ is created by a random flux of charge-$2e$ particles.
 Certainly, the same Fano factor would be seen in a simpler geometry with one tunneling contact. 
The situation is more interesting in the remaining models from Table I. In all of them $q=e$ and the Fano factor in a single-QPC geometry is the same, 
$F=2e$. However, in the two-QPC geometry (Fig. 1), their transport properties are considerably different.

We first consider the Pfaffian state \cite{MR}. It has two edge modes \cite{review}: a charged boson $\phi$ and neutral Majorana fermion $\lambda$
which propagate with different velocities $v_c$ and $v_n$. The Majorana mode contains information about non-Abelian statistics and thus the Pfaffian state exhibits
charge-statistics separation.
The Lagrangian $L_3/\hbar=\int dt dx [-\partial_x\phi(\partial_t+v_c\partial_x)\phi/(2\pi)+
i\lambda(\partial_t+v_n\partial_x)\lambda]$. The electron operator is $\psi=\lambda\exp(-2i\phi)$.
We will need the four-point correlation function for Majorana fermions, $\langle\lambda(1)\lambda(2)\lambda(3)\lambda(4)\rangle=\langle\lambda(1)\lambda(2)\rangle\langle\lambda(3)\lambda(4)\rangle-\langle\lambda(1)\lambda(3)\rangle\langle\lambda(2)\lambda(4)\rangle +\langle\lambda(1)\lambda(4)\rangle\langle\lambda(2)\lambda(3)\rangle $, where
$\langle\lambda(x,t)\lambda(0,0)\rangle=1/[\delta+i(t-x/v_n)]$. The calculations follow the same line as above. The only difference comes from the fact that one needs to take into account the contributions to the current and noise from $\tau{\approx} {-a/v_c}$ (we will denote these contributions $I_c$ and $S_c$) and $\tau\approx -a/v_n$ ($I_n$ and $S_n$).
All contributions $I_{c,n}$ to $I_2$ and $S_{c,n}$ to $S_2$ can be found with the same steps as in the $K=8$ model (one makes a shift $\tau\rightarrow\tau+a/v_{c,n}$ and takes the limit $a\rightarrow\infty$ for the calculation of $I_{c,n}$). One finds $S_c=2eI_c$, $I_n=0$ and $S_n=S_c/4$. Thus, the Fano factor $F_2=(S_c+S_n)/(I_c+I_n)=5e/2$ is universal and exceeds the double carrier charge.

The above result has a simple explanation. After tunneling to edge 3 at QPC1, a charge-$e$ hole splits into
charged and neutral excitations which propagate towards QPC2 with different velocities. When the charged excitations arrives to QPC2,
its energy can be used for the tunneling of the charge $e$ into edge 2.
This process is responsible for $I_c$ and $S_c$.
When a neutral excitation arrives to QPC2 its energy can also be used for a tunneling event.
However, since the creation and annihilation operators of the Majorana fermion $\lambda$ are the same, charge can tunnel both from and to edge 3.
This explains why $I_n=0$. On the other hand, both tunneling directions contribute to the excessive noise $S_n$ and increase
the Fano factor in comparison with a single-mode system.

In the edge-reconstructed Pfaffian state \cite{rec} there are three modes: right-moving charged and neutral Bose-modes 
$\phi_c$ and $\phi_n$ and a left-moving Majorana fermion $\lambda$; 
$L_3/\hbar=1/(4\pi)\int dt dx [-2\partial_x\phi_c(\partial_t+v_c\partial_x)\phi_c-
\partial_x\phi_n(\partial_t+v_n\partial_x)\phi_n + w\partial_x\phi_c\partial_x\phi_n +
4\pi i\lambda(\partial_t-v_\lambda)\lambda]$. Due to the left-moving mode, a non-zero
$S_1$ becomes possible in contrast to the non-reconstructed Pfaffian state.
 The theory has three most relevant electron creation operators on edge 3: $\lambda\exp(2i\phi_c)$ and $\exp(2i\phi_c\pm i\phi_n)$.
Thus, one needs to introduce three pairs of tunneling constants $\Gamma_k^{(\lambda)}$,
$\Gamma_k^{(+)}$ and $\Gamma_k^{(-)}$, where $k=1,2$ labels QPC's.
The interaction between the two Bose-modes affect the Fano factor $F_2$ which depends on all six tunneling constants. We focus instead on the current and noise at QPC1 when $V(S1)=0$, $V(S2)\ne 0$.
Non-zero $S_1$ becomes possible due to the contra-propagating Majorana mode and hence only the tunneling operator $\lambda\exp(2i\phi_c)$ should be taken into account.
The physics and calculation are exactly the same as for $I_n$ and $S_n$ in the Pfaffian case.
We find $I_1=0$; $S_1=\frac{4\pi^3 e^2}{15 \hbar E_c^8}|\Gamma^{(\lambda)}_1\Gamma^{(\lambda)}_2|^2(eV)^5$,
where $E_c$ is the cut-off energy scale of the order of the bulk gap.

We now consider the anti-Pfaffian state \cite{anti}. We start with the simpler non-equilibrated version of that state. It has two contra-propagating charged modes
$\phi_{1,2}$ and a Majorana fermion: 
$L_3/\hbar=1/(4\pi)\int dt dx [-\partial_x\phi_1(\partial_t+v_1\partial_x)\phi_1+
2\partial_x\phi_2(\partial_t-v_2\partial_x)\phi_2+w\partial_x\phi_1
\partial_x\phi_2+4\pi i \lambda(\partial_t-v_{\lambda}\partial_x)\lambda]$.
The model has many independent electron operators but only one most relevant tunneling operator dominates the transport.
Thus, it is sufficient to include only two tunneling constants $\Gamma_{1,2}$ in the model.
Due to the presence of contra-propagating charged modes both $I_1, S_1$ and $I_2, S_2$ are nonzero, if $V(S2)$ or $V(S1)\ne 0$ respectively,
in contrast to all previous models. 
The current and noise $I_1,S_1$ are proportional to $|\Gamma_1\Gamma_2|^2$.
 Hence, the Fano factor $F_1$ is independent of $\Gamma_k$.

The disorder-dominated anti-Pfaffian state \cite{anti} has one charged mode and three contra-propagating 
Majorana modes $\lambda_n$ with the same velocity $v_\lambda$, $L_3/\hbar=
\int dt dx [-\partial_x\phi_c(\partial_t+v_c\partial_x)\phi_c]/(2\pi)+
i\sum_{n=1}^3\lambda_n(\partial_t-v_\lambda\partial_x)\lambda_n]$. 
The tunneling operators become ${(\bf \Gamma}_k{\vec \lambda)}\exp(-2i\phi_c)\epsilon_k^\dagger\eta_k$,
where the tunneling amplitudes ${\bf\Gamma}_k=(\Gamma^{(1)}_k,\Gamma^{(2)}_k,\Gamma^{(3)}_k)$, $k=1,2$, are three-component vectors.
Just like in the non-equilibrated anti-Pfafian model, $I_1,S_1\ne 0$. They are proportional to $\Gamma^4$. Only contributions with two $\Gamma^{(n)}_k$
(with the same or different $n$) and two
complex conjugate $\Gamma^{(m)*}_l$ (with the same or different $m$) are allowed. Each power of $\Gamma_1^{(n)}$ or $\Gamma_1^{(n)*}$
must be accompanied by the same power of $\Gamma_2^{(n)}$ or its conjugate. Besides, the action is invariant with 
respect to orthogonal transformations ${\vec\lambda}\rightarrow\hat O{\vec\lambda}$, 
${\bf\Gamma}_k\rightarrow\hat O{\bf\Gamma}_k$ and hence so are the current and noise. Hence,
$I_2=\frac{2\pi^3 e(eV)^5}{15 \hbar E_c^8}[c_1^I|{\bf\Gamma}_1{\bf\Gamma}_2|^2+c_2^I|{\bf\Gamma}_1{\bf\Gamma}_2^*|^2]$, where $c_l^I$ are constants.
The noise $S_1$ has a similar structure with different constants $c_l^S$ and an overall factor $2e$. 
If only one component of each of the vectors ${\bf\Gamma}_k$
is nonzero then the problem reduces to the edge-reconstructed Pfaffian model. From the comparison with the results for that model one finds
$c_1^I=-c_2^I$, $c_1^S+c_2^S=1$.
Finally, the analysis of the same type as in the $K=8$ model shows that $c_1^S=c_1^I$, $c_2^S=-c_2^I$.
Thus, $F_1=2e[|{\bf\Gamma}_1{\bf\Gamma}_2|^2+|{\bf\Gamma}_1{\bf\Gamma}_2^*|^2]/[|{\bf\Gamma}_1{\bf\Gamma}_2|^2-|{\bf\Gamma}_1{\bf\Gamma}_2^*|^2]$.
From this result one can see a drastic difference between two versions of the anti-Pfaffian model. 
Let a gate electrode modify the shape of edge 3.
This changes the edge disorder contribution to the action. In the non-equilibrated model, it includes terms, linear in $\phi_{1,2}$,
 like $\int dx dt u(x)\partial_x\phi_1$, where $u(x)$ is random.
The disorder contribution to the action of the disorder-dominated model is quadratic in Majorana fermions.
We ignored such contributions so far since they can be gauged out \cite{anti} from the action by a linear transformation of the fields at the expense of changing $\Gamma$'s.
 In the non-equilibrated model this does not affect the Fano factor, independent of $\Gamma_k$.
 The Fano factor depends on the edge disorder and hence the edge shape in the disorder-dominated state \cite{footnote}.

The last state is the Abelian (331) state \cite{ab,rec} with the action
$L_3/\hbar=-1/(4\pi)\int dt dx [3\partial_t\phi_1\partial_x\phi_1-4\partial_t\phi_1\partial_x\phi_2+4\partial_t\phi_2\partial_x\phi_2
+w_{n,m}\partial_x\phi_n\partial_x\phi_m]$. The two most (and equally) relevant electron creation operators 
are $\exp(3i\phi_1-2i\phi_2)$ and $\exp(i\phi_1+2i\phi_2)$. Hence, two pairs of $\Gamma$'s must be included in the model.
Just like in the Pfaffian state, $I_1,S_1=0$. At the same time, one can easily see that $F_2$ depends on the interaction strength $w_{n,m}$ and thus is non-universal in contrast to the Pfaffian and $K=8$ cases.

In conclusion, we suggest an experiment which can distinguish six
candidate states for the 5/2 QHE plateau. The charge-statistics
separation leads to different transport properties in the two-QPC
geometry. The signatures of all states are summarized in Table I. We
acknowledge helpful discussions with B. I. Halperin, M. Heiblum, A.
Pyrkov, B. Rosenow, A. Stern and X.-G. Wen. This work was supported
by NSF grant DMR-0544116 and BSF grant 2006371.

\newpage
\begin{figure}[h]
  \begin{center}
\includegraphics[width=8cm]{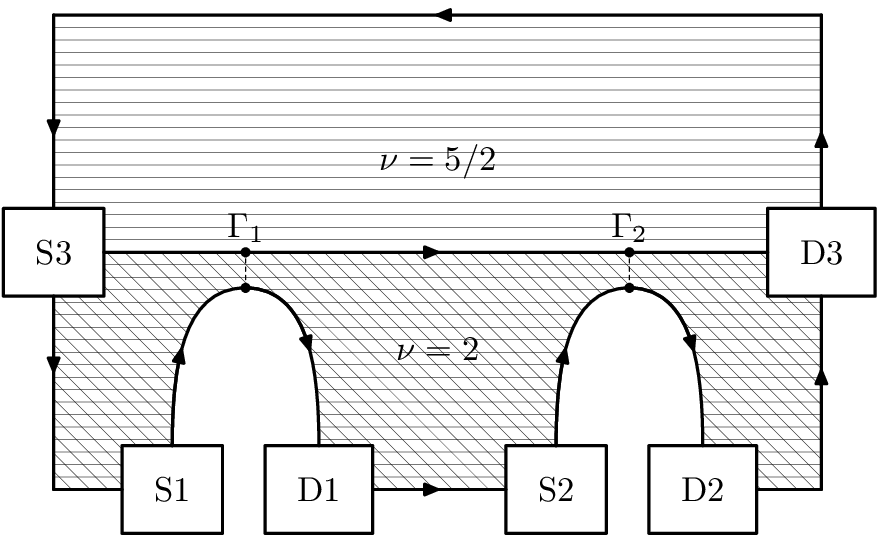}
\end{center}
  \caption{Setup with two quantum point contacts. Arrows show the propagation direction of charged modes.}
\label{fig1}
\end{figure}

\begin{figure}
\begin{center}
\includegraphics[width=8cm,
%angle=270
]{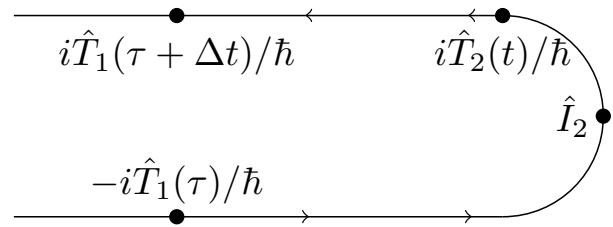}
\end{center}
  \caption{ The Keldysh contour.
    }
\label{fig2}
\end{figure}


\begin{thebibliography} {150}
\bibitem{f0}
The $K=8$ state is formed by tightly bound Cooper pairs and hence electrons are gapped.
\bibitem{footnote}
One can detect the non-equilibrated anti-Pfaffian state from transport measurements in a bar geometry since the quantum Hall conductance $\sigma$ in that state
is $7e^2/(2h)$. Indeed, the left electrode injects charge into the $\phi_1$-channel and two integer QHE edges on the lower edge.
This contributes $3e^2/h$  to $\sigma$. It also injects additional current from the left to right into the
$\phi_2$-channel on the upper edge.
\end{thebibliography}
\end{document}